\documentclass[10pt,twocolumn,a4paper]{article}

\usepackage[T1]{fontenc}
\usepackage{newtxtext,newtxmath}

\usepackage{amsmath}
\usepackage{graphicx}
\usepackage{subcaption}
\usepackage[font=footnotesize]{caption}
\usepackage{threeparttable}
\usepackage{microtype}
\usepackage{xcolor}
\definecolor{blu}{RGB}{7,1,231}
\definecolor{brownish}{RGB}{238,127,22}
\usepackage{bm}
\usepackage{siunitx}
\usepackage{booktabs}
\usepackage{multirow}
\usepackage{adjustbox}
\usepackage[hidelinks]{hyperref}
\usepackage[numbers,sort&compress]{natbib}
\usepackage[a4paper,margin=1.8cm]{geometry}
\newcommand{\keywords}[1]{\begin{center}\textbf{Keywords:} #1\end{center}}

\newcommand{\na}{\textemdash}

\begin{document}

\title{Microscopic Origin of Temperature-Dependent Anisotropic Heat Transport in Ultrawide-Bandgap Rutile GeO$_2$}

\author{
Pouria Emtenani$^{1}$, Marta Loletti$^{2}$, Felix Nippert$^{1}$, Eduardo Bed\^e Barros$^{3,1}$,\\
Zbigniew Galazka$^{4}$, Hans Tornatzky$^{5}$, Christian Thomsen$^{1}$,\\
Juan Sebasti\'an Reparaz$^{2}$, Riccardo Rurali$^{2}$, Markus R. Wagner$^{5,1}$\thanks{Corresponding author: wagner@pdi-berlin.de}
}

\date{}

\maketitle

\noindent
$^{1}$ Technische Universit\"at Berlin, Institute of Physics and Astronomy, Hardenbergstr.~36, 10623 Berlin, Germany\\
$^{2}$ Institut de Ci\`encia de Materials de Barcelona, ICMAB-CSIC, Campus UAB, 08193 Bellaterra, Spain\\
$^{3}$ Department of Physics, Universidade Federal do Cear\'a, Fortaleza, Cear\'a, 60455-760, Brazil\\
$^{4}$ Leibniz-Institut f\"ur Kristallz\"uchtung, Max-Born-Str.~2, 12489 Berlin, Germany\\
$^{5}$ Paul-Drude-Institut f\"ur Festk\"orperelektronik, Leibniz-Institut im Forschungsverbund Berlin e.V., Hausvogteiplatz 5--7, 10117 Berlin, Germany

%AFM abstract
\begin{abstract}
Ultrawide-bandgap rutile GeO$_2$ is emerging as a promising semiconductor for power electronics, where efficient heat dissipation is essential to suppress self-heating and ensure device reliability. However, the temperature dependence and microscopic origin of its anisotropic heat transport have remained experimentally unresolved. Here, temperature-dependent time-domain thermoreflectance measurements combined with first-principles phonon transport calculations are used to quantify the thermal conductivity of single-crystal rutile GeO$_2$ from 80 to 350 K along [001] and [110]. At 295 K, the thermal conductivity reaches 47.5 W\,m$^{-1}$\,K$^{-1}$ along [001] and 32.5 W\,m$^{-1}$\,K$^{-1}$ along [110], corresponding to an anisotropy ratio of 1.46, in good agreement with theory. Rather than following a simple $T^{-1}$ law, the thermal conductivity exhibits an approximate $T^{-1.4}$ dependence, indicating additional scattering beyond purely three-phonon-limited transport. Mode-resolved analysis reveals that the room-temperature anisotropy originates from the combined effect of larger phonon group velocities along [001] and direction-dependent phonon lifetimes. Upon cooling, depopulation of high-frequency phonons progressively suppresses their contribution to heat transport and reduces the anisotropy. The temperature-dependent thermal boundary conductance of Al/rutile GeO$_2$ interfaces is further resolved, and the scaled conductance indicates predominantly elastic interfacial transport. These findings establish the microscopic basis of bulk and interfacial heat transport in rutile GeO$_2$ and position this material as a promising thermally robust platform for ultrawide-bandgap electronics.
\end{abstract}

% ---------- Keywords ----------

\keywords{rutile GeO$_2$, ultra wide bandgap semiconductors, thermal conductivity, anisotropy, power electronics, thermal boundary conductance, time-domain thermoreflectance (TDTR)}
% 
% ---------- Main text ----------

\section{Introduction}

As power electronics advances toward higher efficiency and power density, the demand for semiconductors with exceptional properties has become paramount. The pursuit of energy-efficient current transmission and conversion at elevated capacities while preserving device compactness drives the research community to explore next-generation materials capable of meeting these requirements. The transition from traditional silicon-based semiconductors ($E_g = 1.12$~eV)~\cite{Sze2006} to wide-bandgap materials such as GaN ($E_g = 3.4$~eV)~\cite{miragliotta_transient_1996} and SiC ($E_g = 3.22$~eV)~\cite{wang_multiphoton_2024} has delivered substantial performance gains in power electronics~\cite{rafin_power_2023,roccaforte_emerging_2018}. However, attention has now shifted to ultra-wide-bandgap (UWBG) semiconductors with even larger bandgaps ($E_g > 4$~eV). Among these, $\beta$-Ga$_2$O$_3$ with $E_g \geq 4.48$~eV~\cite{onuma_valence_2015,onuma_temperature-dependent_2016, sturm_dipole_2016,mock_band_2017,meisner_anisotropy_2024} has attracted intense interest due to its theoretical breakdown field of approximately 8 MV/cm and Baliga figure-of-merit of $\sim$3000 (compared to $\sim$800 for GaN and $\sim$300 for SiC)~\cite{baliga_fundamentals_2019,zhang_ultra-wide_2022}, enabling substantially higher power handling capabilities compared to conventional wide-bandgap semiconductors~\cite{xue_overview_2018,higashiwaki_-ga2o3_2022,he_review_2024,woo_wide_2024}.

However, the bandgap alone does not determine the suitability of a material for power electronics: achievable carrier concentration (doping), carrier mobility, and thermal transport are also key figures of merit~\cite{baliga_fundamentals_2019,rowe_computational_2019,chen_accelerating_2025}. Thermal management, in particular, can limit device reliability and power handling because insufficient heat removal leads to self-heating, performance degradation, and premature failure. A prominent example is $\beta$-Ga$_2$O$_3$, whose intrinsically low and strongly anisotropic thermal conductivity has been reported to range from 21.5 to 29.2~W~m$^{-1}$~K$^{-1}$ along [010] and from 9.5 to 16.1~W~m$^{-1}$~K$^{-1}$ along [100]~\cite{guo_anisotropic_2015,jiang_three-dimensional_2018,slomski_anisotropic_2017,klimm_tensor_2023,santia_lattice_2015}. For materials with moderate thermal conductivity, both the bulk thermal conductivity and the thermal boundary conductance of the interfaces can significantly influence junction temperatures~\cite{feng_critical_2023,boteler_thermal_2019,shen_thermal_2017}. Rutile germanium dioxide ($r$-GeO$_2$), with a direct bandgap of approximately 4.64~eV~\cite{stapelbroek_exciton_1978,chae_rutile_2019,mengle_quasiparticle_2019} and substantially higher thermal conductivity than $\beta$-Ga$_2$O$_3$~\cite{chae_thermal_2020,guo_anisotropic_2015}, thus emerges as a compelling UWBG alternative that directly addresses this critical thermal bottleneck.

\begin{figure*}[tbh]
  \centering
    \includegraphics[width=\linewidth]{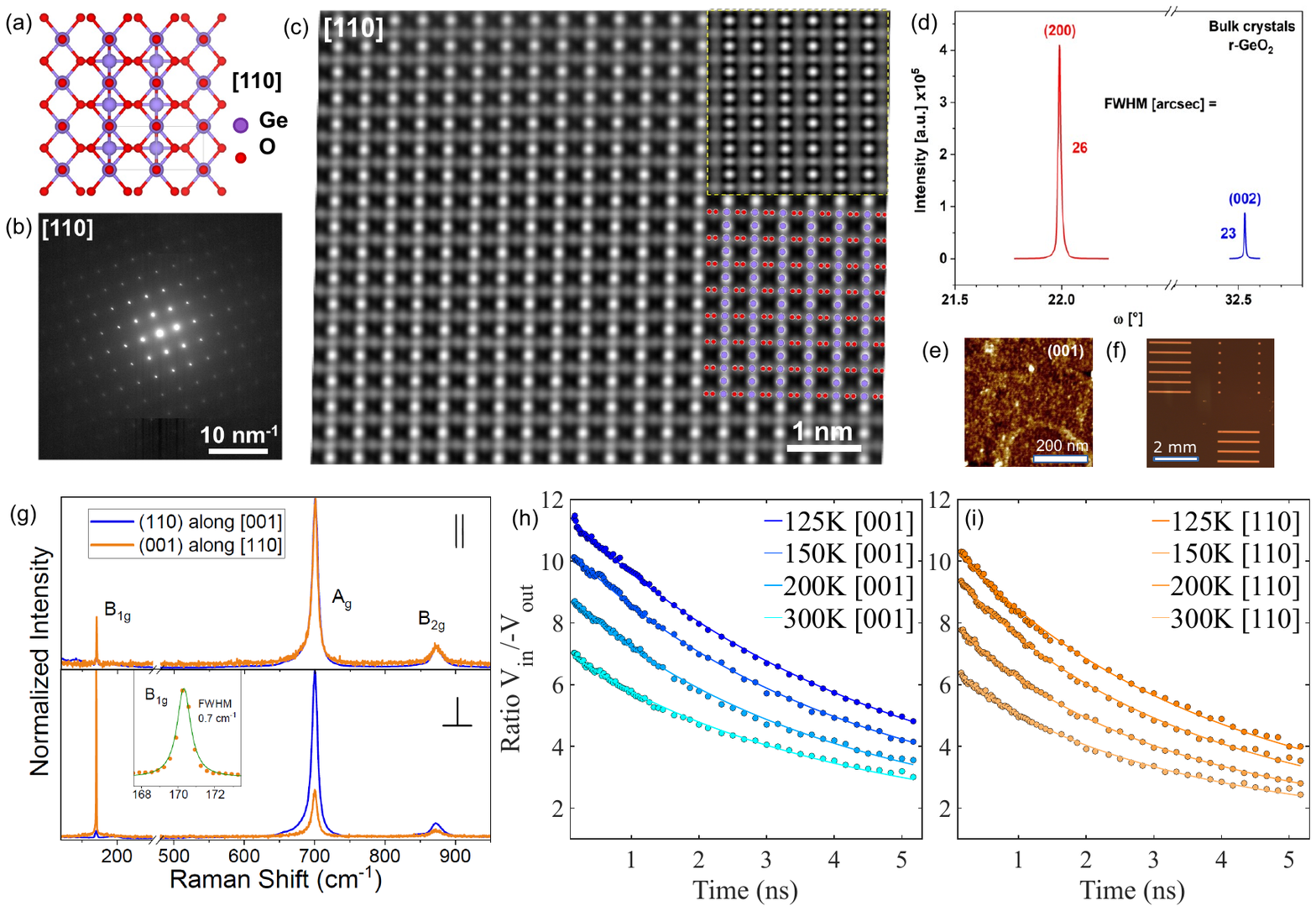}
    \caption{
    properties of \(r\)-GeO\(_2\) samples studied in this work. (a) Schematic illustration of crystal structure in [110] direction, (b) Electron diffraction pattern confirming pure rutile crystal phase, (c) HRTEM image (negative spherical aberration imaging) with insets for simulated HAADF images and superimposed crystal structure, (d) XRD rocking curve of a similar $r$-GeO$_2$ sample, (e) AFM image of (001) plane with RMS roughness of 0.15~nm, (f) optical image of Al transducer mask for TDTR measurements, (g) polarized Raman spectra of (110) and (001) surfaces with incoming polarization along [001] and [110] and outgoing polarization in parallel (upper panel) and crossed (lower panel) configuration, (h, i) Experimental TDTR data (circles) and fitted curves (lines) for the [001] (blue) and [110] (orange) directions. The ratio \(V_{\text{in}} / -V_{\text{out}}\) is shown as a function of pump--probe delay at selected temperatures. Subfigures (a)--(e) are reproduced from Ref.~\cite{galazka_bulk_2025} under the CC BY 4.0 license.}
    \label{fig:fig1ab}
\end{figure*}

Moreover, rutile GeO$_2$ demonstrates theoretical predictions of ambipolar dopability through first-principles defect calculations with identified shallow donor and acceptor states~\cite{chae_rutile_2019,bushick_electron_2020,lyons_ptype_2024}. Recently, Tetzner et al.\ demonstrated successful n-type doping via phosphorus ion implantation, achieving carrier concentrations up to $1.7 \times 10^{19}$~cm$^{-3}$ and enabling the first functional r-GeO$_2$ MOSFETs on single-crystal insulating substrates with breakdown voltages exceeding 1.2~kV~\cite{tetzner_lateral_2025}. Although calculations suggest a suitable valence band position for p-type doping~\cite{niedermeier_shallow_2020}, the realization of efficient p-type conductivity has not yet been demonstrated. The high theoretical electron mobilities (244--377~cm$^2$V$^{-1}$s$^{-1}$ depending on the crystallographic direction)~\cite{bushick_electron_2020} and the feasibility of growing large single crystals by top-seeded solution growth~\cite{galazka_bulk_2025,galazka2026_bulk_rutile_geo2} further strengthen the case for $r$-GeO$_2$ as an emerging UWBG material.

The thermal transport characteristics of $r$-GeO$_2$ have been examined in a very limited number of recent experimental and theoretical studies. Chae et al.\ measured a thermal conductivity of 51~$\pm$~2~W~m$^{-1}$K$^{-1}$ at room temperature for polycrystalline hot-pressed samples and calculated anisotropic single-crystal values of 58~W~m$^{-1}$K$^{-1}$ along [001] and 37~W~m$^{-1}$K$^{-1}$ along [100] using first-principles Boltzmann transport equation (BTE) calculations~\cite{chae_thermal_2020}. More recently, Rahaman et al.\ reported a cross-plane thermal conductivity of 52.9~$\pm$~6.6~W~m$^{-1}$K$^{-1}$ along [001] in MOCVD-grown epitaxial thin films at 300~K~\cite{rahaman_apl_rGeO2_film_2026}. However, to date, there are no experimental works studying the temperature dependent thermal conductivity of $r$-GeO$_2$ along different crystallographic directions in single crystals. Critically, the temperature evolution of thermal anisotropy and its underlying physical mechanisms remain completely unexplored. 

Here, we present the first experimental quantification of the directional thermal conductivity of single-crystal $r$-GeO$_2$ in the temperature range of 80–350 K, thereby revealing a previously unobserved, temperature-dependent anisotropy in its thermal transport behavior. Using a two-color time-domain thermoreflectance (TDTR) setup, we measure the thermal conductivity along the [001] and [110] crystallographic directions of $r$-GeO$_2$ single crystals grown by the top-seeded solution growth method. Coupled with \textit{ab initio} phonon Boltzmann transport equation (BTE) calculations, we not only validate and extend Chae et al.'s theoretical predictions but also uncover the microscopic phonon scattering mechanisms responsible for the temperature-dependent variation in anisotropy. In addition, we determine the temperature-dependent thermal boundary conductance of the Al/$r$-GeO$_2$ interface, providing comprehensive insights into both bulk and interfacial thermal transport physics in this emerging UWBG semiconductor.

\section{Experimental details}

The temperature-dependent thermal transport properties were quantified via time-domain thermoreflectance (TDTR) measurements, enabling determination of the through-plane thermal conductivity of single-crystal rutile germanium dioxide ($r$-GeO$_2$) along the [001] and [110] crystallographic directions. For the measurements, we used samples prepared from bulk $r$-GeO$_2$ single crystals grown by the top-seeded solution growth (TSSG) method at the Leibniz-Institut für Kristallzüchtung. Two samples with lateral dimensions of 5x5mm and thickness of about 0.5~mm with polished (001) and (110) surfaces were prepared from the same single crystal. Figure~\ref{fig:fig1ab} provides an overview of key properties of the $r$-GeO$_2$ single crystals. The pure rutile phase as shown in the structural model in Fig.~\ref{fig:fig1ab}(a) and the excellent structural quality without presence of defects are confirmed by electron diffraction imaging (b) and transmission electron microscopy (c)~\cite{galazka_bulk_2025}. XRD rocking curves of similar $r$-GeO$_2$ samples reveal narrow line widths of diffraction peaks without shoulders (d). The sample surfaces were prepared by chemical mechanical polishing (CMP) yielding a surface roughness of less than 0.2 nm measured by atomic force microscopy (e)~\cite{galazka_bulk_2025}. For TDTR measurements, an Al transducer mask with a thickness of $\sim100$~nm was deposited by electron beam evaporation (f). More details on the growth process and crystal properties of the samples in this study can be found in Galazka et al.~\cite{galazka_bulk_2025, galazka2026_bulk_rutile_geo2}. Raman spectra of the (110) and (001) planes for parallel and crossed polarization reveal the characteristic Raman modes of $r$ -GeO $_2$ with $B_{1g}$, $A_g$ and $B_{2g}$ symmetry (Fig.~\ref{fig:fig1ab}g)~\cite{tornatzky_lattice_2026}. The inset highlights the exceptionally low full-width half-maximum (FWHM) of the $B_{1g}$ mode with a FWHM value of 0.7~cm$^{-1}$. For a full polarization-angle resolved analysis of the Raman modes in $r$-GeO$_2$, see Ref.~\cite{tornatzky_lattice_2026}.

\begin{figure*}[th!]
  \centering
  \begin{subfigure}[t]{0.5\textwidth}
    \centering \vspace{0pt}
    \includegraphics[width=0.9\linewidth]{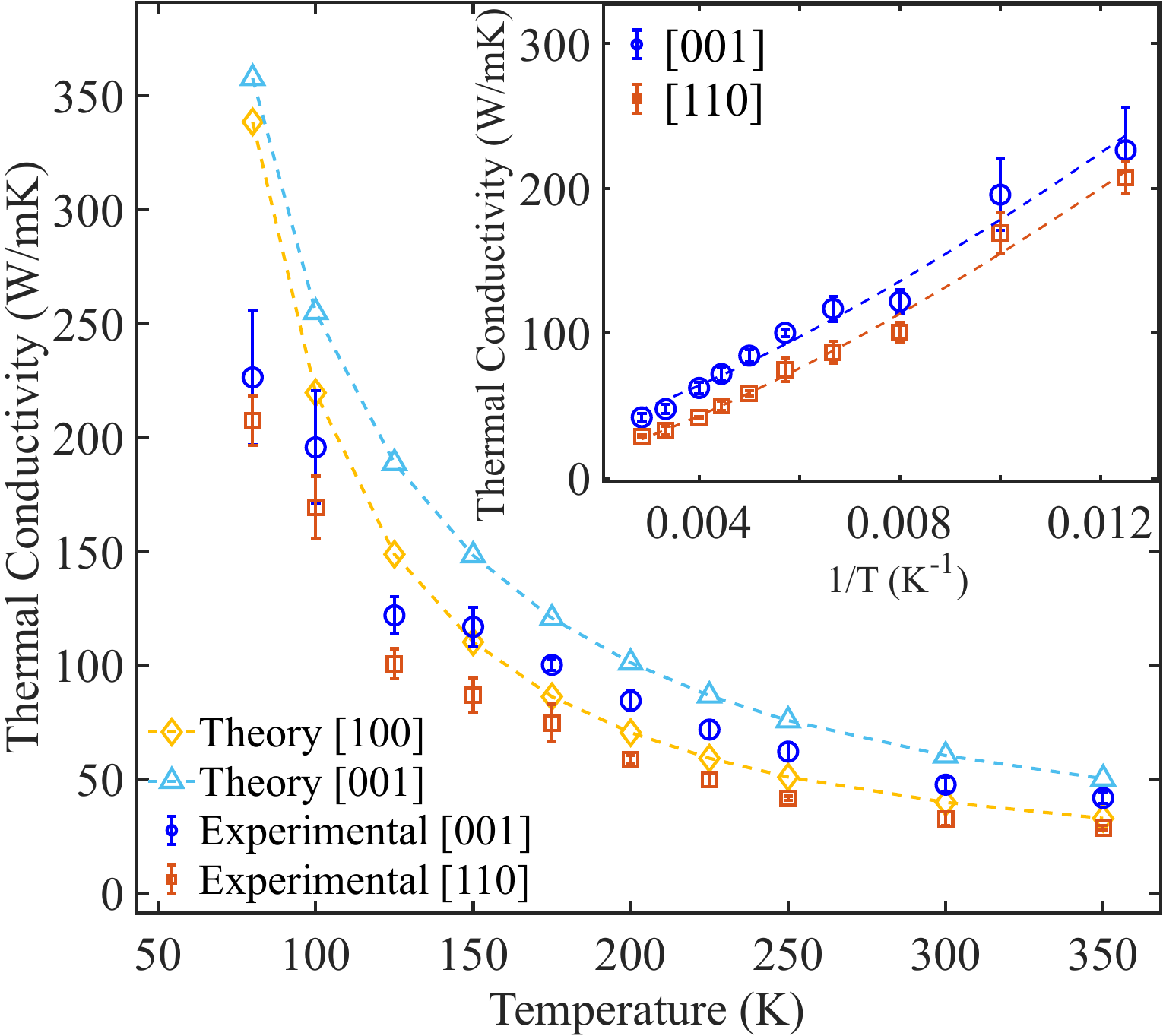}
    \phantomcaption
    \subcaption*{(a)}
    \label{fig:thermal_cond_001_110}
  \end{subfigure}\hfill
  \begin{subfigure}[t]{0.5\textwidth}
    \centering \vspace{0pt}
    \includegraphics[width=0.9\linewidth]{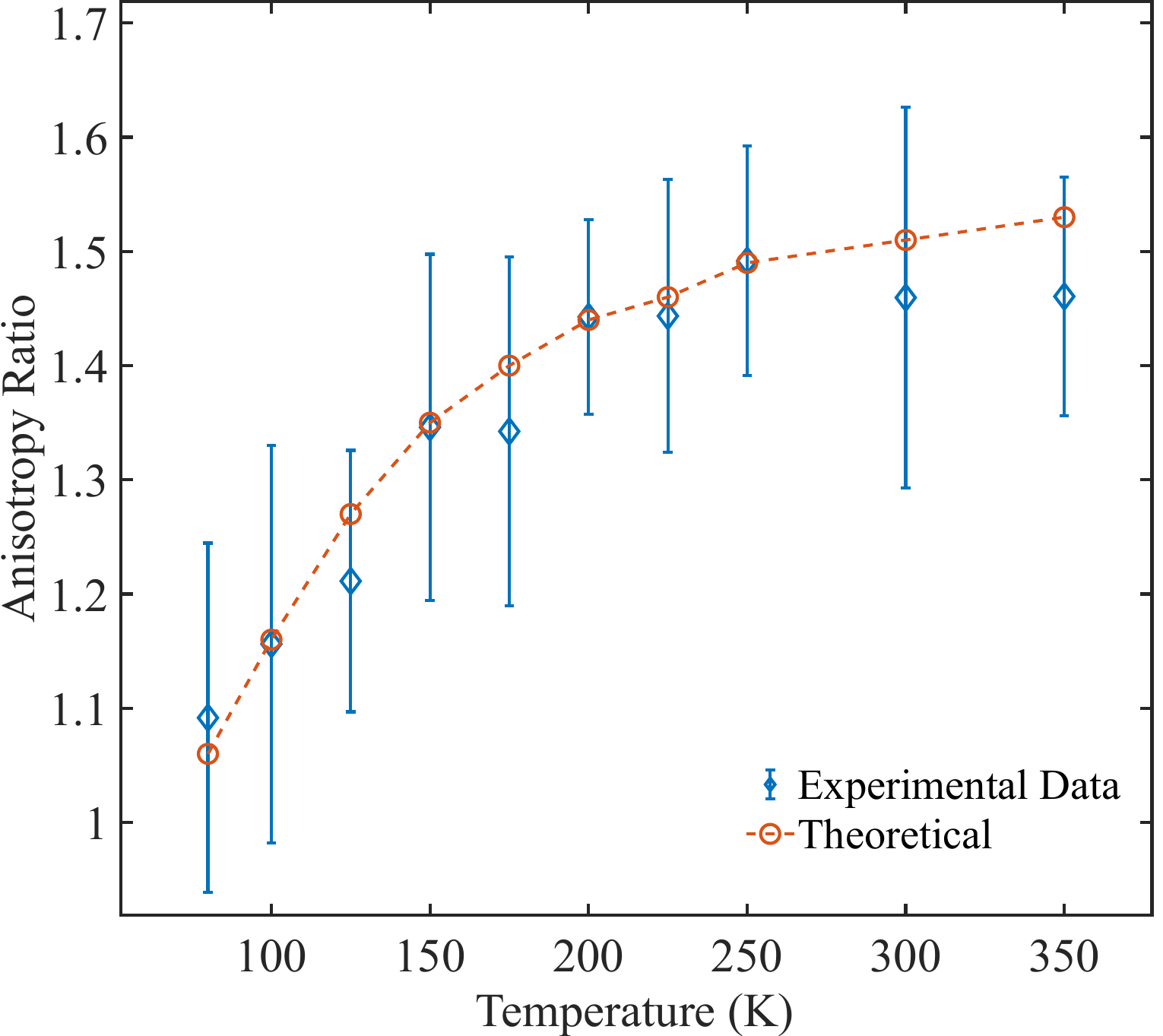}
    \phantomcaption
    \subcaption*{(b)}
    \label{fig:thermal_conductivity_1_over_T}
  \end{subfigure}
  \captionsetup{font=footnotesize}
  \caption{(a) Temperature dependent thermal conductivity $\kappa(T)$ of single-crystal $r$-GeO$_2$ along the [001] and [110] crystallographic directions measured by TDTR (blue and red symbols) and \textit{ab initio} DFT--BTE calculations (light blue and orange symbols and lines). Quantitative thermal conductivity values are listed in Table~\ref{tab:rgeo2_matrix}. Inset: the same experimental data plotted versus $1/T$, highlighting an approximately linear dependency in the displayed temperature range. (b) Anisotropy ratio $\kappa_{[001]}/\kappa_{[110]}$ as a function of temperature. TDTR measurements and DFT--BTE calculations show good agreement and a reduction of anisotropy upon cooling.}
%  \label{fig:tdtr_fit}
\end{figure*}

Figures~\ref{fig:fig1ab}(h) and (i) display the measured and fitted TDTR data for the crystallographic directions [001] and [110] measured on the (001) and (110) planes, respectively. Each panel includes four temperature datasets for the corresponding orientation. The solid circles represent the experimental data points for the in-phase to out-of-phase signal ratio \((-V_{\text{in}} / V_{\text{out}})\) as a function of the time delay between pump and probe pulses. The solid lines correspond to the fitted curves obtained using a diffusive thermal transport model introduced by Cahill~\cite{cahill_analysis_2004}, with simulations implemented following Refs.~\cite{FeserTDTR, ThohenseeTDTR}. The volumetric heat capacity of $r$-GeO$_2$ was obtained from first-principles density-functional theory (DFT) phonon calculations and compared with the literature values~\cite{counsell_entropy_1967}~(Figure~S3).
The heat capacity and thermal conductivity of Al were taken from the literature~\cite{hust_thermal_1984,ditmars_aluminum_1985}.

A clear increase of the ratio \(R = V_{\mathrm{in}}/{-}V_{\mathrm{out}}\) is observed with decreasing temperature. This behavior can be attributed to the fact that, at fixed modulation frequency \(f\), the out-of-phase amplitude follows
\[
|V_{\mathrm{out}}| \propto (k\,C\,f)^{-1/2} = \frac{1}{e\,\sqrt{f}}, \quad e = \sqrt{k\,C},
\]
with \(e\) the thermal effusivity \cite{cahill_analysis_2004}.
On cooling, the rise of thermal conductivity \(k(T)\) typically outweighs the modest drop of heat capacity \(C(T)\),
so \(e\) increases, \(|V_{\mathrm{out}}|\) decreases, and the ratio \(R=-V_{\mathrm{in}}/V_{\mathrm{out}}\) correspondingly rises \cite{koh_frequency_2007}.

\section{Results and Discussion}

\begin{figure*}[th]
    \centering
    \includegraphics[width=1 \textwidth]{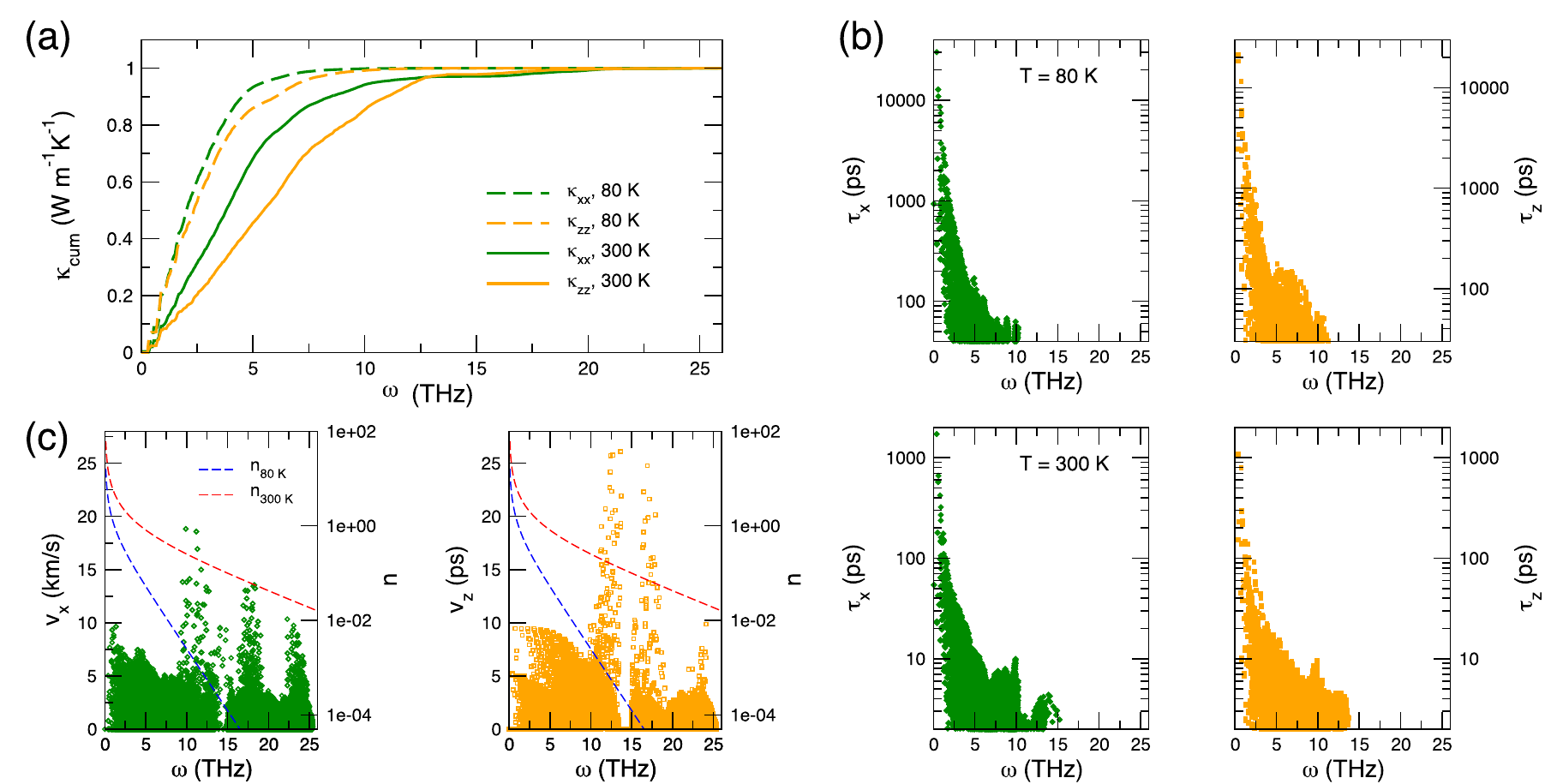}
   \caption{(a) Spectrally accumulated (cumulative) thermal conductivity $\kappa_{\mathrm{cum}}(\omega)$ along the $z$ ([001]) and $x$ ([110]) directions at $T=80$~K (dashed) and $T=300$~K (solid), plotted as a function of phonon frequency $\omega$. The accumulation is shown in normalized form (saturating to unity) to emphasize the frequency range contributing to heat transport in each direction.
(b) Frequency-dependent phonon lifetimes for heat flow along $x$ and $z$, $\tau_x(\omega)$ and $\tau_z(\omega)$, at $T=80$~K (top row) and $T=300$~K (bottom row); we define $\tau_i = \tau \frac{v_i}{|v|}$ as the lifetime of phonons propagating along the $i$-axis.
(c) Phonon group velocities as a function of frequency for transport along $x$ and $z$, $v_x(\omega)$ (left) and $v_z(\omega)$ (right). Dashed curves indicate the equilibrium Bose--Einstein occupation factor $n_0(\omega,T)$ at 80 and 300~K (right axis), illustrating the progressive depopulation of high-frequency modes upon cooling and its impact on the spectral contributions to $\kappa$.}
    \label{fig:calc}
\end{figure*}

Figure~\ref{fig:thermal_cond_001_110} shows the thermal conductivity of $r$-GeO$_2$ single crystals measured along the crystallographic directions [110] ($\kappa_{xx}$) and [001] ($\kappa_{zz}$) in the temperature range between 80~K to 350~K. Due to the tetragonal symmetry of rutile-type structures, the thermal conductivity tensor has only two independent components, $\kappa_{xx} = \kappa_{yy}$ and $\kappa_{zz}$. As a result, the thermal conductivity along the [100] and [110] directions is equal~\cite{ding_anisotropic_2014}. For room temperature (295~K), we experimentally quantify the anisotropic thermal conductivity of $r$-GeO$_2$ as $\kappa_{xx}=32.5\pm3.0~\mathrm{W\,m^{-1}\,K^{-1}}$ and $\kappa_{zz}=47.5\pm3.2~\mathrm{W\,m^{-1}\,K^{-1}}$ (see Table~\ref{tab:rgeo2_matrix} for T dependent values).

The inset of Figure~\ref{fig:thermal_cond_001_110} displays the same values of the measured thermal conductivity of $r$-GeO$_2$ plotted versus \(1/T\) for both the [001] and [110] directions. The temperature range of our measurements, from 80~K to 350~K, corresponds to approximately 0.1 to 0.5 times the Debye temperature of \(r\)-GeO\(_2\), with reported values between 687~K and 767~K~\cite{liu_first-principles_2010, hermet_thermodynamic_2013}. According to thermal transport theory~\cite{grimvall1999}, the maximum thermal conductivity typically occurs near 0.1\(\Theta_D\), whereas thermal conductivity is expected to scale inversely with temperature (\(k \sim 1/T\)) due to three-phonon scattering processes in the range of 0.1–0.5\(\Theta_D\). Within the investigated temperature range (80--350\,K), our experimental data are better described by an effective $\sim 1/T^{1.4}$ dependence than by a simple $1/T$ trend. Such an effective exponent can reflect the combined influence of multiple scattering mechanisms and anharmonic effects (including thermal expansion and higher-order phonon--phonon scattering beyond the simplest three-phonon-limited picture); for example, four-phonon scattering has been shown to reduce lattice thermal conductivity in a range of solids~\cite{grimvall1999, feng_four_2017}.

The experimental data for both directions in Figure~\ref{fig:thermal_cond_001_110} are compared to our \textit{ab initio} calculations based on density-functional theory (DFT) and solution of the Boltzmann transport equation (BTE) beyond the relaxation time approximation (RTA) (c.f. Methods section 5.2). The agreement between our theoretical results, similar to those previously reported by Chae et al.~\cite{chae_thermal_2020}, and our experimental measurements is very good. Both theory and experiments reveal a very similar temperature dependence and highlight the anisotropic behavior of $r$-GeO$_2$, with a higher thermal conductivity in the [001] direction (i.e., $\kappa_{zz} > \kappa_{xx}$). Comparing the absolute values, we find that the measured thermal conductivities are systematically lower than the \textit{ab initio} predictions, with an offset of approximately 20\% at room temperature. This discrepancy is consistent with observations in other wide-bandgap semiconductors such as GaN, SiC, AlN, and $\beta$-Ga$_2$O$_3$, where measured values typically fall 15--30\% below theoretical predictions for intrinsic three-phonon scattering~\cite{zheng_thermal_2019,xu_thermal_2019,vaca_measurements_2022}. The difference arises as calculations model an idealized defect-free crystal, whereas real samples contain additional scattering centers including point defects such as vacancies or interstitials, impurities, and extended structural defects~\cite{scott_phonon_2018,rounds_influence_2018}. Despite the offset in absolute values, the excellent agreement in the temperature dependence and anisotropy ratios confirms that the dominant phonon scattering mechanisms are correctly captured in our calculations.

In the following discussion, we focus on the temperature dependent anisotropy of the thermal conductivity in $r$-GeO$_2$ and elucidate its microscopic origin. Figure~\ref{fig:thermal_conductivity_1_over_T} illustrates the thermal conductivity anisotropy ratio $\kappa_{zz}/\kappa_{xx}$ in the temperature range studied. At room-temperature, we find an experimentally determined anisotropy ratio of 1.46 in very good agreement with the ratio of 1.51 as derived by \textit{ab initio} calculations. The anisotropy ratio exhibits a monotonic decrease with decreasing temperature, indicating a progressive reduction in the disparity between the thermal conductivities along the [001] and [110] crystallographic directions.

In principle, the observed temperature evolution can arise from two coupled effects: (i) a temperature-driven change in lattice stiffness, which modifies phonon group velocities, and (ii) the progressive suppression of phonon--phonon scattering upon cooling, which increases phonon relaxation times. To disentangle these contributions, we analyzed direction-resolved spectra along the $x$ and $z$ axes, as summarized in Figure~\ref{fig:calc}. We first consider the spectral accumulation of the thermal conductivity. Figure~\ref{fig:calc}a shows the cumulative conductivity $\kappa_{\mathrm{cum}}(\omega)$ versus phonon frequency $\omega$ for both axes at 300~K and 80~K. At 300~K, $\kappa_{\mathrm{cum}}(\omega)$ exhibits a pronounced directional dependence, whereas at 80~K the two curves display a much more similar frequency evolution. This trend mirrors our experimental observation that the thermal anisotropy diminishes upon cooling. Notably, the characteristic frequency at which $\kappa_{\mathrm{cum}}$ approaches its high-frequency saturation (i.e., where the remaining modes contribute negligibly) shifts for a T reduction from 300~K to 80~K from $\approx12.5$~THz to $\approx8$~THz for $\kappa_{zz}$ and from $\approx10$~THz to $\approx6$~THz for $\kappa_{xx}$. This difference can be rationalized by examining phonon group velocities and relaxation times in each direction. 

Our calculations indicate that at room temperature both phonon group velocities and phonon lifetimes contribute to the anisotropy, which is reflected in the distinct spectral evolution of $\kappa_{\mathrm{cum}}$ along the two axes. Because heat transport is frequency dependent, the relative weight of these mechanisms varies across the spectrum; nevertheless, a clear overall trend emerges. Specifically, we find that on average the phonon velocities along $z$ are systematically larger than along $x$, i.e., $v_z > v_x$. Thus, phonons propagating in $z$ direction exhibit higher characteristic velocities than those traveling perpendicular to z. This behavior is captured by Figure~\ref{fig:calc}c, which reports the spectral dependence of the group velocities in both directions and shows a persistent velocity advantage along $z$ over a broad frequency range.

\begin{figure}[thb]
    \centering
    \includegraphics[width=0.48\textwidth]{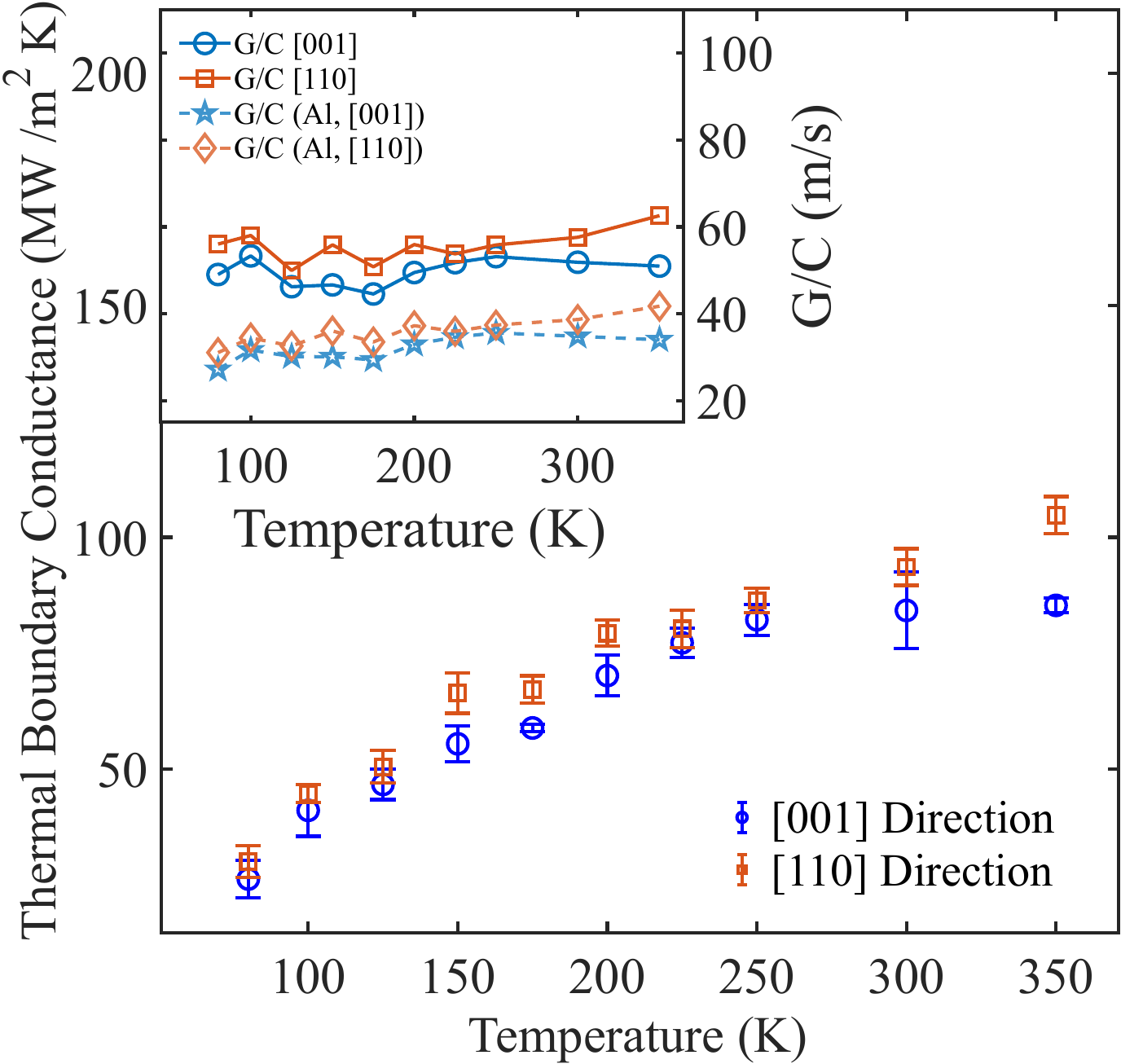}
    \caption{Interfacial thermal boundary conductance $G(T)$ for Al/r-GeO$_2$ in the [001] and [110] orientations obtained from the multilayer TDTR fits. Inset: Corresponding scaled quantities $G/C_v^{\le 10\,\mathrm{THz}}$ and $G/C_{\mathrm{Al}}$, showing an almost temperature-independent behaviour.}
    \label{fig:tbc}
\end{figure}

At the same time, the room-temperature relaxation times also favor anisotropic heat transport. Figure~\ref{fig:calc}b shows the spectral dependence of the phonon relaxation times for modes contributing to thermal transport in each direction. Although the lifetimes are comparable over most of the spectrum, the 300~K data indicate a pronounced reduction of $\tau$ for phonons traveling along the $x$-axis within the $\sim$10--12.5 THz phonon frequency window. This selective suppression reduces the contribution of high-frequency phonons to $\kappa$ along the $x$ direction, precisely in the frequency range where $\kappa_{\mathrm{cum}}(\omega)$ continues to accumulate along $z$ at 300~K, thereby triggering a substantial difference between $\kappa$ along both directions. Taken together, the combination of larger group velocities and larger effective lifetimes along $z$ at higher frequencies provides a consistent microscopic basis for the observed thermal-conductivity anisotropy between the $x$ and $z$ directions at room temperature.

As temperature decreases, the role of phonon group velocities in setting the thermal-conductivity anisotropy progressively diminishes. To understand the origin of this effect, it is useful to revisit Figure~\ref{fig:calc}a, where we plot the cumulative (spectrally accumulated) thermal conductivity as a function of frequency. In this representation, $\kappa_{\mathrm{cum}}(\bar{\omega})$ denotes the thermal conductivity obtained by including only phonon modes with $\omega < \bar{\omega}$. At $T=300$~K, the accumulation continues well into the high-frequency portion of the spectrum, with phonons up to $\sim$12.5~THz still providing a non-negligible contribution to ${\kappa}$. In contrast, at 80~K the cumulative curves in both directions saturate much earlier, reaching convergence already around $\sim$6--8~THz. This marked reduction of high-frequency contributions upon cooling is consistent with the Bose--Einstein statistics: as shown in Figure~\ref{fig:calc}c, the occupation of phonons with $\omega \gtrsim 10$~THz becomes negligible at low temperature, such that these modes contribute little to heat transport. Consequently, the anisotropy mechanism identified at room temperature (where high-frequency modes along $z$ combine relatively large group velocities with direction-dependent relaxation times) is progressively quenched as those modes are depopulated. Phonon lifetimes further support this picture: upon cooling, the spectral distribution of $\tau$ does not exhibit a pronounced directional anomaly in the same high-frequency window (10--12.5~THz) as discussed for 300~K. Taken together, the depletion of phonon populations in the high-frequency region (characterized by features around $\sim$12.5~THz at 300~K) suppresses the portion of the spectrum in which the $z$-axis velocity advantage is most efficiently converted into a larger thermal conductivity. This mechanism accounts for the experimentally observed suppression of thermal anisotropy at low temperatures.

Furthermore, we extract the interfacial thermal boundary conductance, \(G\), from the same multilayer TDTR fits used to determine the substrate thermal conductivity. The resulting temperature dependence \(G(T)\) is summarized in Figure~\ref{fig:tbc} for interfaces with the crystal planes (001) and (110). In both cases, \(G\) decreases monotonically upon cooling from 350 to 80~K, consistent with a progressive reduction in the population of heat-carrying phonons available to participate in interfacial energy exchange. The two orientations exhibit comparable magnitudes and similar trends across the entire temperature range, indicating that any crystallographic dependence of the effective transmission is relatively modest compared with the overall population-driven temperature dependence visible in the raw \(G(T)\) curves.

\begin{table*}[thb]
  \centering
  \small
  \setlength{\tabcolsep}{3.5pt}
  \renewcommand{\arraystretch}{1.08}

  \begin{threeparttable}
   \caption{Temperature-dependent thermal conductivity along the [001] ($\kappa_{zz}$) and [110] ($\kappa_{xx}$) crystallographic directions, together with the thermal anisotropy ratio $\kappa_{zz}/\kappa_{xx}$. Results obtain is this work are listed in comparison to reported experimental and computational values in the literature.}
    \label{tab:rgeo2_matrix}

    \begin{tabular*}{\textwidth}{@{\extracolsep{\fill}} lcccccccccccc}
      \toprule
      \multirow{3}{*}{T (K)} &
      \multicolumn{4}{c}{$\kappa_{zz}$ (\si{W.m^{-1}.K^{-1}})} &
      \multicolumn{3}{c}{$\kappa_{xx}$ (\si{W.m^{-1}.K^{-1}})} &
      $\kappa_{\mathrm{avg}}$ &
      \multicolumn{3}{c}{Anisotropy $\kappa_{zz}/\kappa_{xx}$} \\
      \cmidrule(lr){2-5} \cmidrule(lr){6-8} \cmidrule(lr){9-9} \cmidrule(lr){10-12}
      & \multicolumn{2}{c}{This work} & Chae~\cite{chae_thermal_2020} & Rahaman~\cite{rahaman_apl_rGeO2_film_2026} &
        \multicolumn{2}{c}{This work} & Chae~\cite{chae_thermal_2020} &
        Chae~\cite{chae_thermal_2020} &
        \multicolumn{2}{c}{This work} & Chae~\cite{chae_thermal_2020} \\
      & TDTR & DFT+BTE & DFT+BTE$^{a}$ & TDTR$^{b}$ &
        TDTR & DFT+BTE & DFT+BTE$^{a}$ &
        laser flash$^{c}$ &
        TDTR & DFT+BTE & DFT+BTE$^{a}$ \\
      \midrule
       80 & 226.3 & 357.7 & \na & \na & 207.3 & 338.5 & \na & \na & 1.09 & 1.06 & \na \\
      100 & 195.6 & 255.0 & 255.0 & \na & 169.2 & 219.6 & 195.4 & \na & 1.16 & 1.16 & 1.31 \\
      125 & 121.9 & 188.7 & \na & \na & 100.6 & 148.7 & \na & \na & 1.21 & 1.27 & \na \\
      150 & 116.8 & 148.2 & 145.5 & \na & 86.8 & 110.2 & 99.2 & \na & 1.35 & 1.34 & 1.47 \\
      175 & 100.1 & 120.6 & \na & \na & 74.6 & 86.2 & \na & \na & 1.34 & 1.40 & \na \\
      200 & 84.4 & 101.1 & 98.5 & \na & 58.5 & 70.4 & 63.6 & \na & 1.44 & 1.44 & 1.55 \\
      225 & 71.7 & 86.6 & \na & \na & 49.7 & 59.2 & \na & \na & 1.44 & 1.46 & \na \\
      250 & 62.1 & 75.7 & 73.4 & \na & 41.6 & 50.9 & 46.1 & \na & 1.49 & 1.49 & 1.59 \\
      295 & 47.6 & 60.3 & 58.4 & 52.9 & 32.6 & 39.8 & 36.1 & 51.0 & 1.46 & 1.52 & 1.62 \\
      350 & 41.7 & 50.2 & 48.5 & \na & 28.6 & 32.8 & 29.7 & \na & 1.46 & 1.53 & 1.63 \\

% --- Laser-flash temps with interpolated Chae single-crystal DFT+BTE^a (avoid long 400--1000 K block) ---
376 & \na & \na & 44.0 & \na & \na & \na & 27.3 & 44.8 & \na & \na & 1.61 \\
478 & \na & \na & 34.1 & \na & \na & \na & 20.6 & 34.0 & \na & \na & 1.65 \\
596 & \na & \na & 26.5 & \na & \na & \na & 15.9 & 29.0 & \na & \na & 1.66 \\
677 & \na & \na & 23.4 & \na & \na & \na & 14.0 & 24.3 & \na & \na & 1.67 \\
 
      \bottomrule
    \end{tabular*}

    \begin{tablenotes}[flushleft]
      \footnotesize
      \item[\tnote{a}] Chae et al.\ \cite{chae_thermal_2020}: single-crystal DFT/BTE calculations (LDA, \textsc{almaBTE}).
      \item[\tnote{b}] Rahaman et al.\ \cite{rahaman_apl_rGeO2_film_2026}: reports only RT value along $[002]$; for rutile, $[002]\parallel[001]$.
      \item[\tnote{c}] Chae et al.\ \cite{chae_thermal_2020}: hot-pressed polycrystalline r-GeO$_2$ bulk measured by laser-flash calorimetry; orientation-averaged $\kappa_{\mathrm{avg}}$.
      \item[] Missing entries are marked by \na.
      \item[] This work: TDTR measurements and DFT+BTE calculations with LDA and \textsc{almaBTE}.
    \end{tablenotes}
  \end{threeparttable}
\end{table*}

To separate phonon population from transmission effects at the interface, the inset in Figure~\ref{fig:tbc} re-plots the same data in terms of scaled conductances. We normalize \(G\) by \(C_v^{\le 10\,\mathrm{THz}}\), the r-GeO$_2$ lattice heat capacity accumulated up to \(\sim 10\)~THz (a practical cutoff set by the highest-frequency phonons in Al that can participate efficiently in elastic transmission) and, independently, by the volumetric heat capacity of the Al transducer, \(C_{\mathrm{Al}}\)~\cite{cheng_thermal_2020}. Remarkably, both scaled quantities, \(G/C_v^{\le 10\,\mathrm{THz}}\) and \(G/C_{\mathrm{Al}}\), remain approximately temperature independent from 80 to 350~K for both orientations. This near-constant behavior implies that the dominant temperature dependence in \(G(T)\) arises primarily from the thermal occupation (and thus the available interfacial energy flux) rather than from a strongly temperature-dependent transmission probability. In other words, once the phonon population is accounted for (either on the r-GeO$_2$ side within the Al spectral window or on the Al side via \(C_{\mathrm{Al}}\)), the effective interfacial coupling is nearly constant, supporting a picture in which heat transfer is governed largely by elastic phonon processes and constrained by the \(\sim 10\)~THz spectral window of Al. This scaling collapse provides additional evidence that inelastic scattering channels— which would ordinarily generate supplementary temperature dependence beyond that dictated by simple thermal occupation (population) factors—do not constitute the dominant contribution within the investigated temperature interval. 

\section{Conclusions}
We combined TDTR measurements and \textit{ab initio} phonon-transport calculations to clarify anisotropic heat conduction in single-crystal rutile GeO$_2$ in the temperature range between 80--350~K. The measured thermal conductivities $\kappa_{zz}$ along [001] and $\kappa_{xx}$ along [110] agree closely with DFT-based solutions of the phonon Boltzmann transport equation beyond the relaxation-time approximation, confirming a higher conductivity along the tetragonal axis, $\kappa_{zz}>\kappa_{xx}=\kappa_{yy}$. At room temperature, the anisotropy ratio is $\kappa_{zz}/\kappa_{xx}\approx 1.46$ from TDTR, in good agreement with the calculated value of 1.51. Within the investigated temperature range, the data are best described by an approximate $1/T^{1.4}$ dependence rather than by a simple $1/T$ trend, indicating contributions beyond purely three-phonon-limited transport.

Mode-resolved analysis shows that the room-temperature anisotropy arises from the combined effect of larger phonon group velocities along [001] and direction-dependent phonon lifetimes within a high-frequency spectral window that still contributes appreciably at 300~K. Upon cooling, Bose--Einstein depopulation strongly suppresses these high-frequency contributions and the cumulative thermal conductivity saturates at substantially lower frequencies, explaining the observed reduction and eventual collapse of the anisotropy at low temperature. Finally, we extracted the Al/$r$-GeO$_2$ thermal boundary conductance for [001] and [110] interfaces. While \(G\) decreases with decreasing temperature, scaling by the heat capacities of the transmitting phonon populations yields nearly temperature-independent ratios, consistent with largely elastic interfacial transport constrained by the $\sim$10~THz spectral window of Al. Together, these results provide a unified experimental and first-principles understanding of bulk anisotropy and interfacial heat transfer in rutile GeO$_2$, relevant to thermal management and device integration.

\section{Methods}
\label{sec:Methods}

\subsection{Time-Domain Thermoreflectance}
\label{sec:Methods-TDTR}
Time-domain thermoreflectance (TDTR) measurements are conducted in a home-built setup which is schematically illustrated in Fig.~S1 of the Supplementary Information. Our implementation is based on a two-color pump-probe scheme. We employed a low-noise fiber laser (NKT Photonics / APE Emerald Engine Duo longpulse) system that delivers dual synchronized output beams at 1030 nm and its second harmonic (SHG) at 515 nm, serving as the probe and pump beams, respectively. The laser operates at a 76 MHz repetition rate with a pulse width of 5.5 ps. The pump beam is first directed through a resonant electro-optic modulator (Qubic AM2B-VIS\_10) with a modulation frequency of 10 MHz, which is directly driven by the Zurich Instruments lock-in amplifier (HF2LI), then travels over a one-meter delay stage (Newport IMS100LM-SA), providing decay times of more than 6.6 ns. After passing through the delay stage, the beam is coupled into a single-mode fiber, which is attached to the triplet fiber coupler (TC18FC-532). This arrangement greatly enhances the alignment precision and significantly minimizes beam divergence. The single-mode fiber directs the beam to a cold mirror, which then reflects it into a Mitutoyo objective (M Plan Apo NIR 20X). The cold mirror also serves to block back reflections from the sample, enhancing measurement accuracy. The probe beam is directly coupled into a single-mode fiber attached to the triplet fiber coupler (TC18FC-1030), which helps maintain a Gaussian beam profile and contributes to the compactness of the setup. After exiting the single-mode fiber, the beam is directed towards the objective using a 30:70 (R:T) beam splitter to maximize the collection of back-reflected light from the sample, which is then routed into the avalanche Photodetector(Thorlabs APD431C). The pump and probe beams are then focused onto the sample through the 20X Mitutoyo Apochromat objective. 

The samples are mounted in a liquid helium / nitrogen flow cryostat (Janis ST-500), allowing temperature-dependent measurements with temperature stability of \(\pm\)50 mK. Spatial mapping is performed by a two-stage system comprising heavy-duty $x$-$y$ linear positioning stages with long travel range (PI M-413.1DG) to support the cryostat, while precision translation is achieved by a multi-axis piezo stage (PI Nanocube P-611.3 XYZ). Lastly, any residual leakage of the pump beam is effectively blocked by a longpass color filter placed in front of the detector.

Aluminum transducer masks of nominal 100~nm thickness were deposited by electron-beam evaporation (Pfeiffer Classic 500). The transducer thickness was measured in-situ by the quartz crystal microbalance method and confirmed by a profilometer (Bruker DektakXT). The pump- and probe beam diameters at the sample plane were measured using the reflection knife-edge method, yielding \((1/e^{2})\) diameters of \(6.5\,\mu\mathrm{m}\) (pump) and \(3.8\,\mu\mathrm{m}\) (probe). Considering that the pump and probe spot sizes are significantly larger than the thermal penetration depth at the used modulation frequency of 10~MHz, we model heat flow as predominantly one-dimensional (normal to the surface) following standard TDTR practice, such that the fitted conductivity corresponds primarily to the cross-plane component~\cite{jiang_tutorial_2018}.

To confirm the reliability of our TDTR measurements, we used a GaAs wafer as a reference and measured its temperature-dependent thermal conductivity, comparing the data with published literature values~\cite{luo_gallium_2013}. The heat capacity of GaAs was taken from Ref~\cite{passler_representative_2011}. Our results show excellent agreement with published data across the measured temperature range (Supplementary Information, Fig.~S2).

\subsection{Raman Spectroscopy}
Raman spectra were measured in a Horiba LabRam Evolution spectrometer using the 632.8\,nm line of a HeNe laser, a 100x objective (NA=0.9), an 1800 l/mm grating, and with the monochromator-entrance aperture set to 20\,µm, to achieve best spectral resolution. In this configuration the spectral resolution is determined to be better than 0.5\,cm$^{-1}$, as determined from the FWHM of the emission of a low-pressure Ne-lamp.
\subsection{Calculations}
\label{sec:Methods-Calc}

We performed density functional theory (DFT) calculations with the VASP package~\cite{KressePRB93, Kresse94, KressePRB99}, using an energy cutoff of 600~eV, the projector augmented wave method~\cite{blochl_projector_1994}, and the local density approximation (LDA)~\cite{PerdewZungerPRB81, CeperleyPRL80} for the exchange-correlation energy. The Brillouin zone was sampled with a grid of $6 \times 6 \times 9$ {\bf k}-points. With these settings we obtained lattice parameters of the tetragonal structure $a=b=4.3823$~\AA\ and $c = 2.8729$~\AA. We computed the second- and third-order interatomic force constants (IFCs) by finite differences using phonopy~\cite{TogoPRB15} and thirdorder.py~\cite{LiCPC14} in a $5 \times 5 \times 7$ and a $4 \times 4 \times 6$ supercell, respectively; the phonon dispersion is displayed in Figure S4 of the SI. We also computed the second-order IFCs in supercells up to $6 \times 6 \times 9$ with the pre-trained force-field MACE~\cite{Batatia2025DesignSpaceE3Equivariant, Batatia2022mace}, verifying that a $5 \times 5 \times 7$ supercell already provides very well converged phonon dispersions. The calculated phonon frequencies at the $\Gamma$-point are in good agreement with recent experimental polarization-resolved Raman spectroscopy measurements on $r$-GeO$_2$ single crystals~\cite{tornatzky_lattice_2026}, validating our density-functional perturbation theory approach.

The IFCs were used to solve the Boltzmann Transport Equation (BTE) beyond the Relaxation Time Approximation (RTA) on a grid of $28 \times 28 \times 42$ {\bf q}-points with the almaBTE~\cite{CarreteCPC17} code. The lattice thermal conductivity, within the relaxation time approximation (RTA), has the following expression:

\begin{equation}
\kappa_{ij} = \sum_\lambda c_\lambda v_{i,\lambda}v_{j,\lambda} \tau_\lambda
,
\label{eq:rta}
\end{equation}
where $i$ and $j$ are the spatial directions $x$, $y$, and $z$, and the sum runs over all phonon modes, the index $\lambda$ including both ${\bf q}$-point and phonon band. In Eq.~\ref{eq:rta},
$c_\lambda$, ${\bm v}_{\lambda}$, and $\tau_\lambda$ are, respectively, the volumetric heat capacity, group velocity and lifetime of phonon $\lambda$.

The full solution of the BTE employed here reads:
\begin{equation}
\kappa_{ij} = \sum_\lambda c_\lambda v_{i,\lambda}F_{j,\lambda}
,
\label{eq:full}
\end{equation}
where ${\bm F}_{\lambda}$ is a generalized mean free path~\cite{LiCPC14}. Isotopic scattering, using the natural abundances of isotopes of Ge and O, was considered by means of the model of Tamura~\cite{TamuraPRB83}.

\section*{Acknowledgments}
This work was funded by the Leibniz-Gemeinschaft (Senatsausschuss Wettbewerb -- SAW, Germany) under grant no.\ K417/2021. It was partly carried out within the framework of GraFOx, a Leibniz ScienceCampus partially funded by the Leibniz Association.
M. L., J. S. R., and R. R. acknowledge financial support by MCIN/AEI/10.13039/501100011033 under grant PDC2023-145934-I00, PID2024-162811NB-I00, and the Severo Ochoa Centres of
Excellence Program under grant CEX2023-001263-S. M. L. has received funding from the European Union’s Horizon Europe research and innovation programme under the Marie Skłodowska-Curie grant agreement No 10181337 (DOCFAM+). We thank the Centro de Supercomputaci\'on de Galicia (CESGA) for the use of their computational resources.
The authors thank Roberts Blukis from Leibniz-Institut für Kristallzüchtung for critical reading of the paper.

%\section*{CRediT authorship contribution statement}
% Put actual roles here before submission.

\section*{Data availability statement}
All data supporting the findings of this study are available within the article and its Supplementary Information. Additional raw TDTR data and first-principles calculation outputs are available on reasonable request.

\section*{Declaration of competing interest}
The authors declare no competing financial interests.

\bibliographystyle{unsrtnat}
\bibliography{reference}

\clearpage
\onecolumn
\appendix
\setcounter{figure}{0}
\renewcommand{\thefigure}{S\arabic{figure}}
\setcounter{table}{0}
\renewcommand{\thetable}{S\arabic{table}}

\section*{Supplementary Information}
\subsection*{Time-Domain Thermoreflectance Setup}

\begin{figure*}[htb]
\centering
\includegraphics[width=0.70\textwidth]{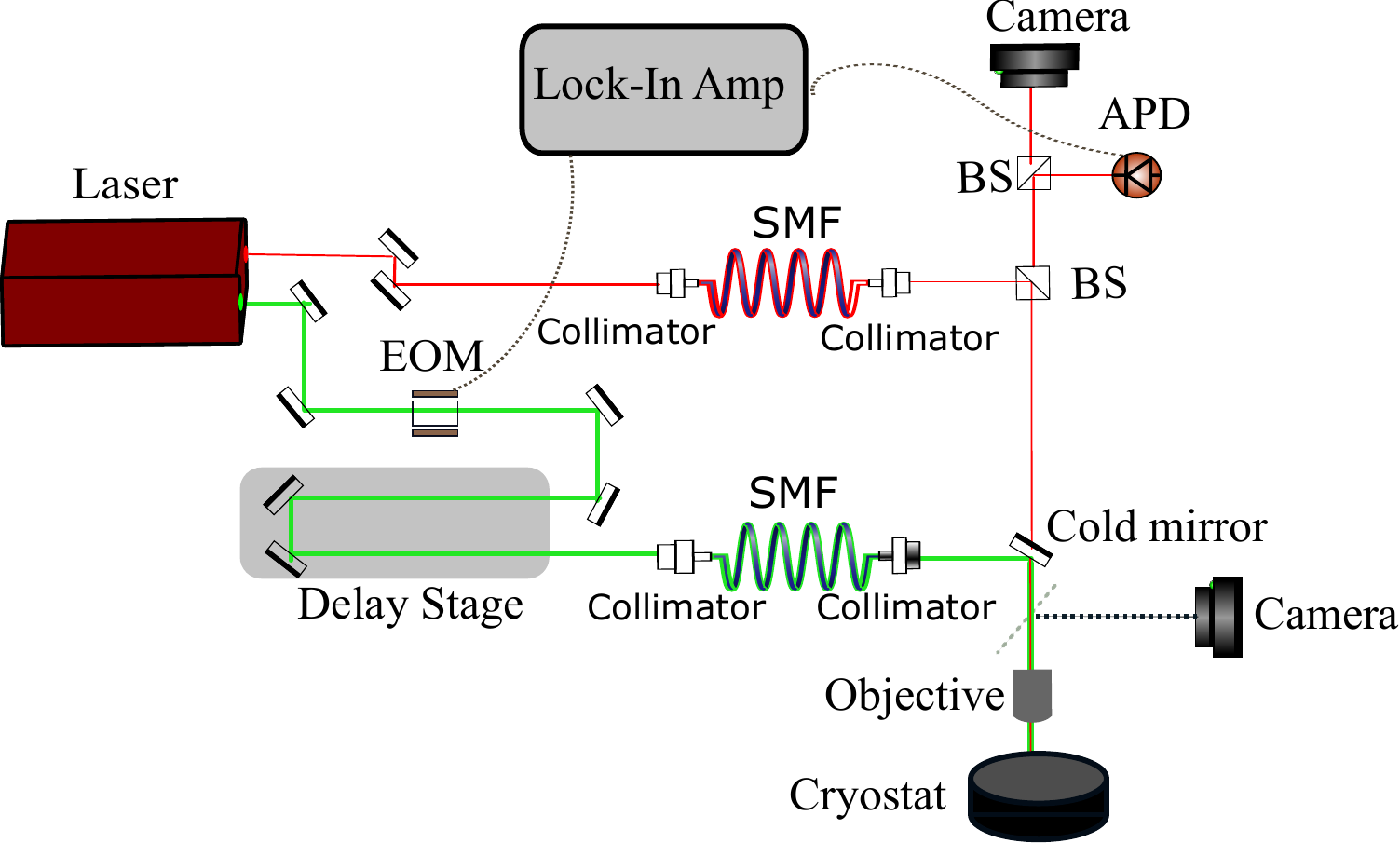}
\caption{(a) Schematic of the experimental setup: The green line represents the path of the pump beam (515 nm), which is directed through the electro-optic modulator (EOM) and then to the delay stage. After the delay stage, the beam is coupled into a single-mode fiber (SMF) and is then guided through a cold mirror to the objective. The probe beam (1030 nm), indicated by a red line, is coupled into the SMF. Subsequent to its passage through the SMF, the beam is reflected by the beam splitter to the objective. Two cameras are employed to monitor the alignment of the beams and their positions on the sample during the course of the measurements. Back reflections from the pump are blocked by the cold mirror and a long-pass colour filter positioned in front of the avalanche Photodetector(APD), which is directly connected to the lock-in amplifier (LIA).}
\label{fig:tdtr_setup}
\end{figure*}

\subsection*{Reflection Knife-Edge Method}

To determine the pump and probe spot sizes directly at the measurement plane, we employed a reflection knife-edge method implemented on the same samples used for TDTR. During Al transducer deposition, we applied optical photolithography with a $5''$ chrome-on-quartz mask to define rectangular Al pads with sharp edges on the same substrates. These Al--substrate transitions provide a high-reflectivity contrast and act as knife edges.

The pump (or probe) beam was focused onto the sample with the same objective and working distance as in TDTR, and the sample was translated laterally with a multi-axis piezo stage to drive the Al edge across the beam with nanometre precision. The back-reflected power was recorded continuously while scanning. The resulting S-shaped transition (from mostly bare to fully on Al) was fitted with the improved knife-edge model of de Araújo \textit{et al.}~\cite{S_araujo}
, which uses a calibrated logistic approximation to the Gaussian error-function edge response. From this fit we consistently extracted the beam radius (reported as the $1/e^{2}$ intensity radius) across all scans.

This procedure was applied to both pump and probe on the actual TDTR samples, and the resulting spot sizes were used in the thermal model without further correction.

\subsection*{TDTR Reference Measurement of GaAs}

\begin{figure}[h!]
  \centering
  \includegraphics[width=0.50\textwidth]{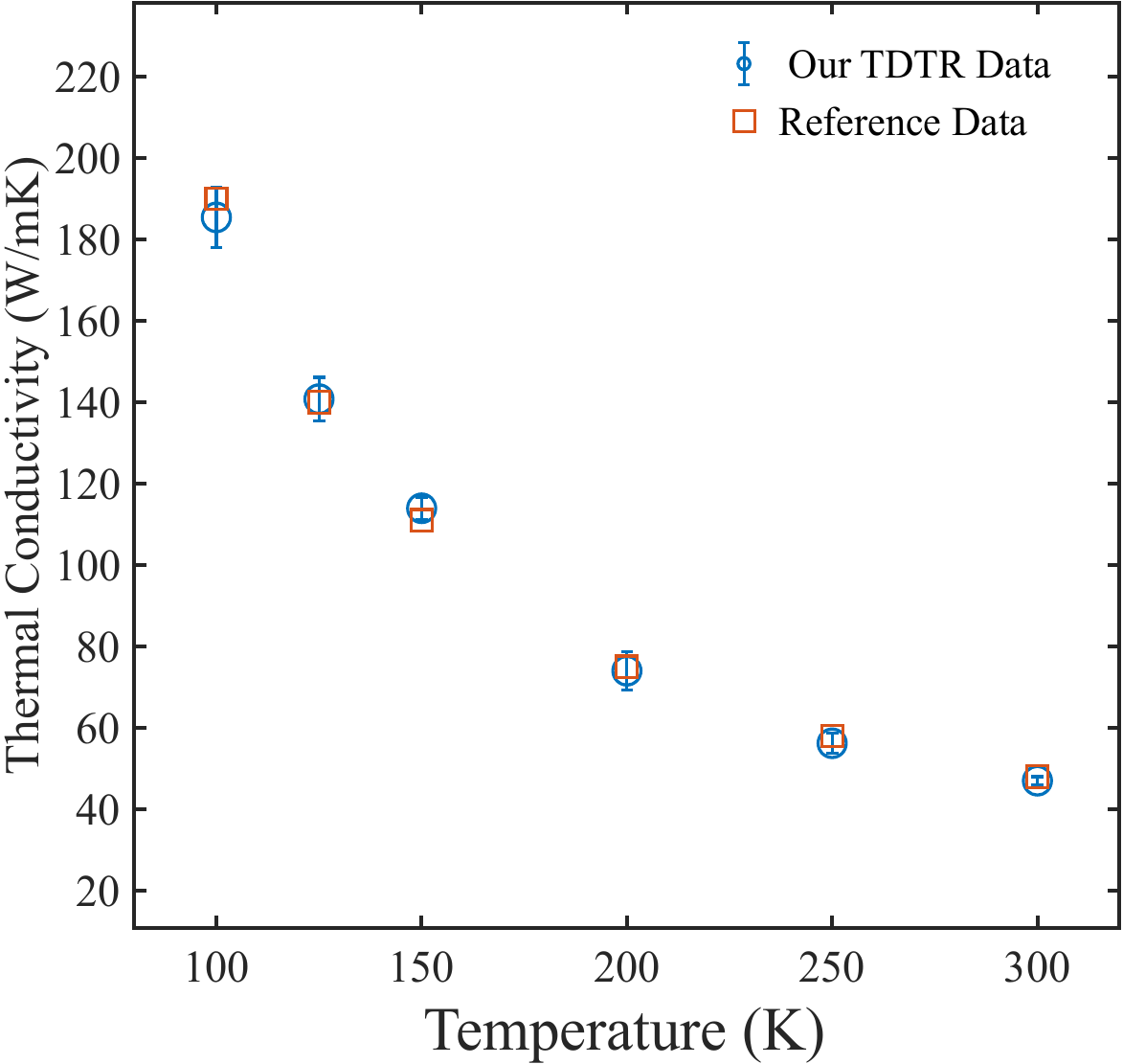}
    \caption{Reference measurement on GaAs: measured \(k(T)\) (blue circles) compared with literature data~\cite{S_luo} (orange squares).}
    \label{fig:GaAs}
\end{figure}

Figure~\ref{fig:GaAs} illustrates the thermal conductivity of bulk GaAs measured with our TDTR setup and used as a reference sample. The temperature-dependent thermal conductivity was compared with data from the literature \cite{S_luo} For the analysis, the heat capacity of GaAs was taken from~\cite{S_passler}. The measured values are in excellent agreement with the reported data over the entire measured temperature range.

\clearpage

\subsection*{Heat Capacity of $r$-GeO$_2$}

\begin{figure}[h!]
  \centering
  \includegraphics[width=0.50\linewidth]{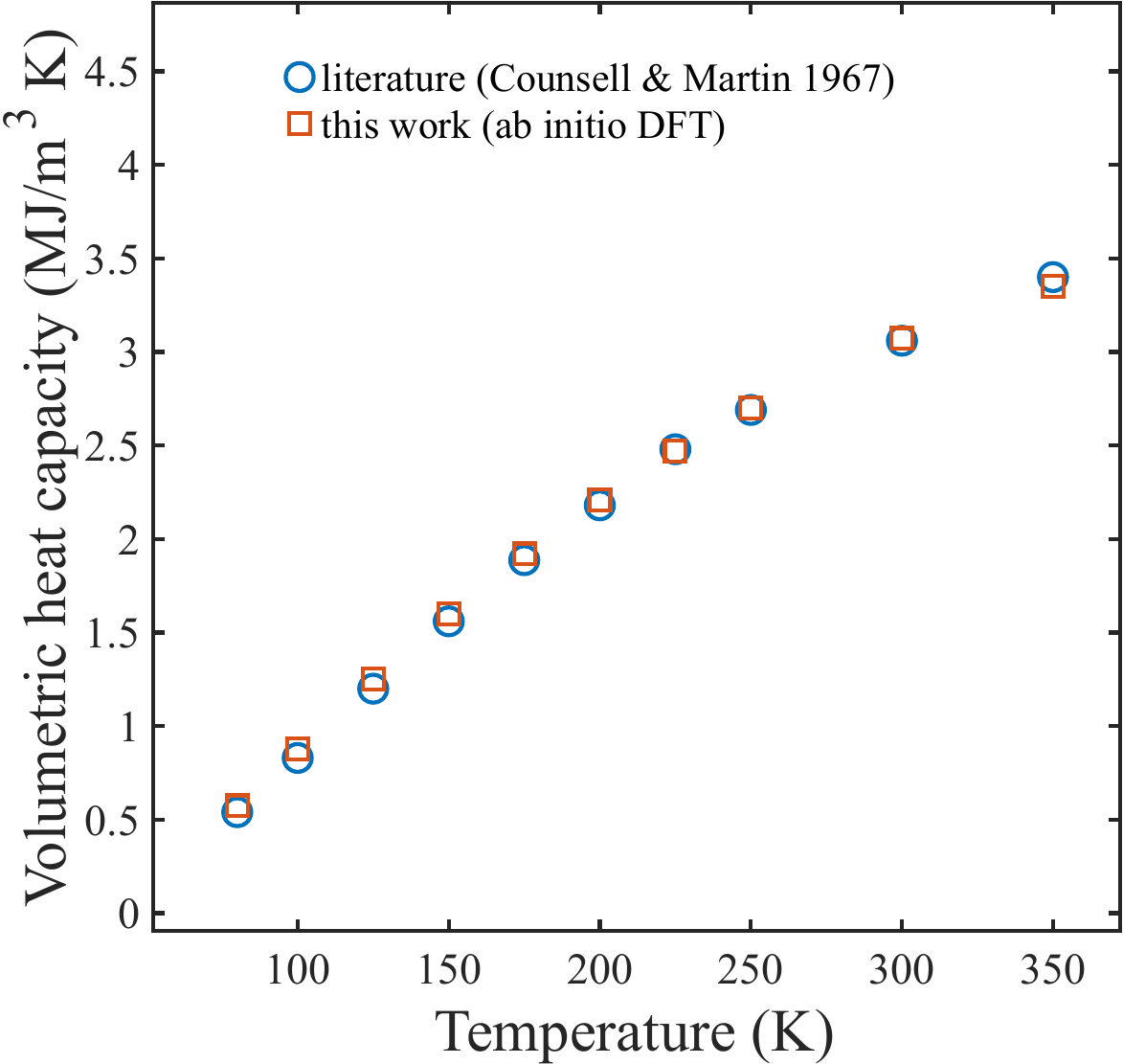}
\caption{Volumetric heat capacity of $r$-GeO$_2$. Blue circles: DFT phonon calculation. Orange squares: literature data from Counsell \& Martin~\cite{S_counsell}, (converted to volumetric units).}
\label{fig:rGeO2_Cv}
\end{figure}

\subsection*{Phonon Dispersion Relation of $r$-GeO$_2$}

\begin{figure}[h!]
  \centering
  \includegraphics[width=0.70\linewidth]{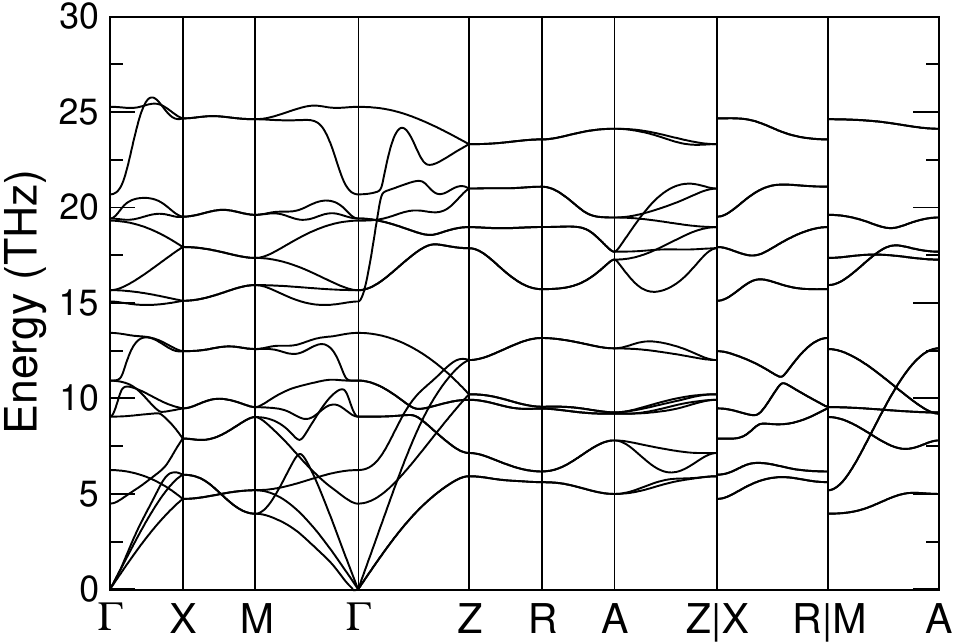}
\caption{Calculated phonon dispersion of $r$-GeO$_2$.}
\label{fig:dispersion}
\end{figure}

\clearpage

\section*{References for Supplementary Information}

\end{document}